# ANOMALOUS DIFFUSION OF EPICENTRES

Oscar Sotolongo-Costa (1,2), R. Gamez (1,2), A. Posadas (2,3,4), F, Luzón(3,4)

*(1) Departamento de Física Teórica. Universidad de la Habana. Cuba.*
*(2) Cátedra de Sistemas Complejos "Henri Poincaré", Facultad de Física, Universidad de la Habana, Cuba.*
*(3) Departamento de Física Aplicada. Universidad de Almería. Spain.*
*(4) Instituto Andaluz de Geofísica. Granada. Spain.*

## *ABSTRACT*

The classification of earthquakes in main shocks and aftershocks by a method recently proposed by M. Baiesi and M. Paczuski allows to the generation of a complex network composed of clusters that group the most correlated events.
The spatial distribution of epicentres inside these structures corresponding to the catalogue of earthquakes in the eastern region of Cuba shows anomalous anti-diffusive behaviour evidencing the attractive nature of the main shock and the possible description in terms of fractional kinetics.

## *I. - INTRODUCTION*

The diffusive behaviour of earthquakes has been a topic of interest in the last years. Several works in this direction use a continuous time random walk (CTRW) model to explain the motion of the epicentres [1-4].
The possibility to introduce a description in terms of  diffusion emerges if it is analyzed in main shock-aftershock structures. In the present paper we show that the spatial distribution of events inside main shock-aftershock cells (MAC) obtained by the Baiesi and Paczuski (BP) correlation method [5, 6] can be described in terms of diffusion process. A MAC is defined by the principal event, the main shock, and all the other events that can be correlated with it by the BP model.
As will be seen, the introduction of such a new classification of earthquakes in mainshocks and aftershocks leads to new and interesting results that may reveal subjacent fractional dynamics, characteristic of anomalous diffusion processes, and a very curious effect of anti-diffusion, apparently due to some sort of "size effect" of the MACs
Usually, the classification of main shocks and aftershocks is made by the space-time windows method. The aftershocks are identified by counting all the events within a space time window [5, 6] but this method does not define the probability that collected event is correlated with the main event. Another problem is the estimation of the size of the space-time window.
The BP method shows a seismic network in which the distinction between main shocks and aftershocks is relative. This kind of network (BP network) reveals properties very similar to those already known in the theory of dynamic networks and, inside it, a representative MAC can be defined if the number of aftershocks correlated with the main shock is relatively high.
In the present paper we study the diffusive behaviour of epicentres inside MACs with many events. In this way a description of the diffusion of the epicentres by the fractional kinetics is shown.

So, the migration of epicentres inside each MAC was analyzed, resulting that epicentre diffusion behaves as a subdiffusive CTRW but with a temporal evolution that resembles a process of concentration of the aftershocks towards the point where the main shock originated.
Section II is devoted to the presentation of the essentials of the BP method, while in Section III we explain how the BP method was used to process the catalogue of earthquakes from the eastern section of Cuba
In Section IV some aspects of anomalous diffusion and the fractional calculus to describe it are presented., and in Section V the results of this analysis are presented, showing a confirmation of the results obtained in [5,6], adding the description of the earthquake distribution with concepts borrowed from the theory of anomalous diffusion. In Section VI we try to explain the curious anti-diffusive behaviour here observed in terms of size effect of the MACs determined by the BP method.

## *II.-BP method*

The arbitrariness in the classification of earthquakes in mainshocks and aftershocks is a still unsolved problem for which the method proposed in [5-7] seems to us a good quantitative criterion to describe the complex correlation of earthquakes in space, time and magnitude.
The usual approach to classify aftershocks is by counting all events within a fixed and to some extent arbitrary space-time window following a pre-assigned main event. This method does not give the probability that an event thereby collected is actually correlated to the main event.
In [5], a more objective criterion to classify earthquakes is introduced through a metric based in the fact that the average number of earthquakes within an interval $\Delta m$ of the magnitude $m$ in a seismic area of fractal dimension $d_f$ and size $r$ over a time interval $t$ is:

$$\bar{n} = Ctr^{d_f} \Delta m 10^{-bm} ,$$

being $C$ an adjustment constant depending on the region, and $b \approx 1$ is the Gutenberg-Richter constant.
Consequently, given an event $j$, occurring at time $t_j$ and looking backward in time to events $i$ occurring at times $t_i < t_j$, the correlation between the two events that occur in a seismic region is defined as $\gamma_{ij} = n_{ij}^{-1}$

where

$$n_{ij} \equiv Ctl_{ij}^{d_f} \Delta m 10^{-bm_i} ,$$

being now $l_{ij}$ the spatial distance between the events and $t = t_j - t_i$.

Of all the earthquakes preceding $j$, the most unlike to occur (and then the more correlated, since it, in fact, occurs) is that for which $n_{ij}$ is minimal.

Then, the degree of correlation between two earthquakes can be quantitatively decided, and is inversely proportional to $n_{ij}$.

When we apply this method to all the earthquakes, and link those events for which $\gamma_{ij}$ maximizes, hierarchical structures (networks) can emerge, in which the bigger event in the cluster is a "hub", called the main event, but also some aftershocks can appear, revealing that the classification in main shocks and aftershocks is less restrictive with this method, and no event is *a priori* purely an aftershock or a main shock. Both emerge as limiting cases of a broad spectrum of related events, where every one can be a precursor or an aftershock. Note that $n_{ij}$ is here the weight of the links. Then, event $j$ is considered to be a replica of event $i$ if $n_{ij}$ is smaller than a previously given $n_c$. The values of $n_{ij}$ too large indicate that the event $j$ is weakly correlated with the previous $i$ events. By removing all "weak" links ($n_{ij} > n_c$), the network is decomposed into clusters. The correlated events are reliably detected when $n_c$ is smaller than one but not too small, since in this case a very fragmented network appears due to detachment of the correlated events. On the other hand, for large $n_c$, some weakly correlated events make links, generating a giant cluster. In this method those earthquakes with magnitude smaller than a previously given $m_<$ are discarded.

## III. - Construction of main shock-aftershock cells (MACs)

Here, we apply the BP criterion to construct the MACs for the catalogue of earthquakes of the south eastern region of Cuba [8]. If, considering a pair of events $i$ and $j$, it is found that $n_{ij} \ll n_c$, then the correlation between them is high. Here we choose $n_c = 10^{-4}$ and $\Delta m = 1$. Besides, only events with threshold magnitude $m_< = 1.5$ were selected.

For each event $j$, the $i$-th event of all previous ones for which $n_{ij}$ takes the minimum value was chosen. The MACs were so determined and their area was calculated in all cases. The distance between earthquakes was calculated transforming the longitude and latitude to plane coordinates. The $d_f$ was determined by the box counting method getting the value $d_f = 1.4$. The value of $b$ was obtained through the Gutenberg-Richter law getting $b = 0.94 \pm 0.03$. The value of the constant $C$ is determined by the procedure given in [1] and we found $C \cong 10^{-8}$. Fig. 1 shows the studied region.

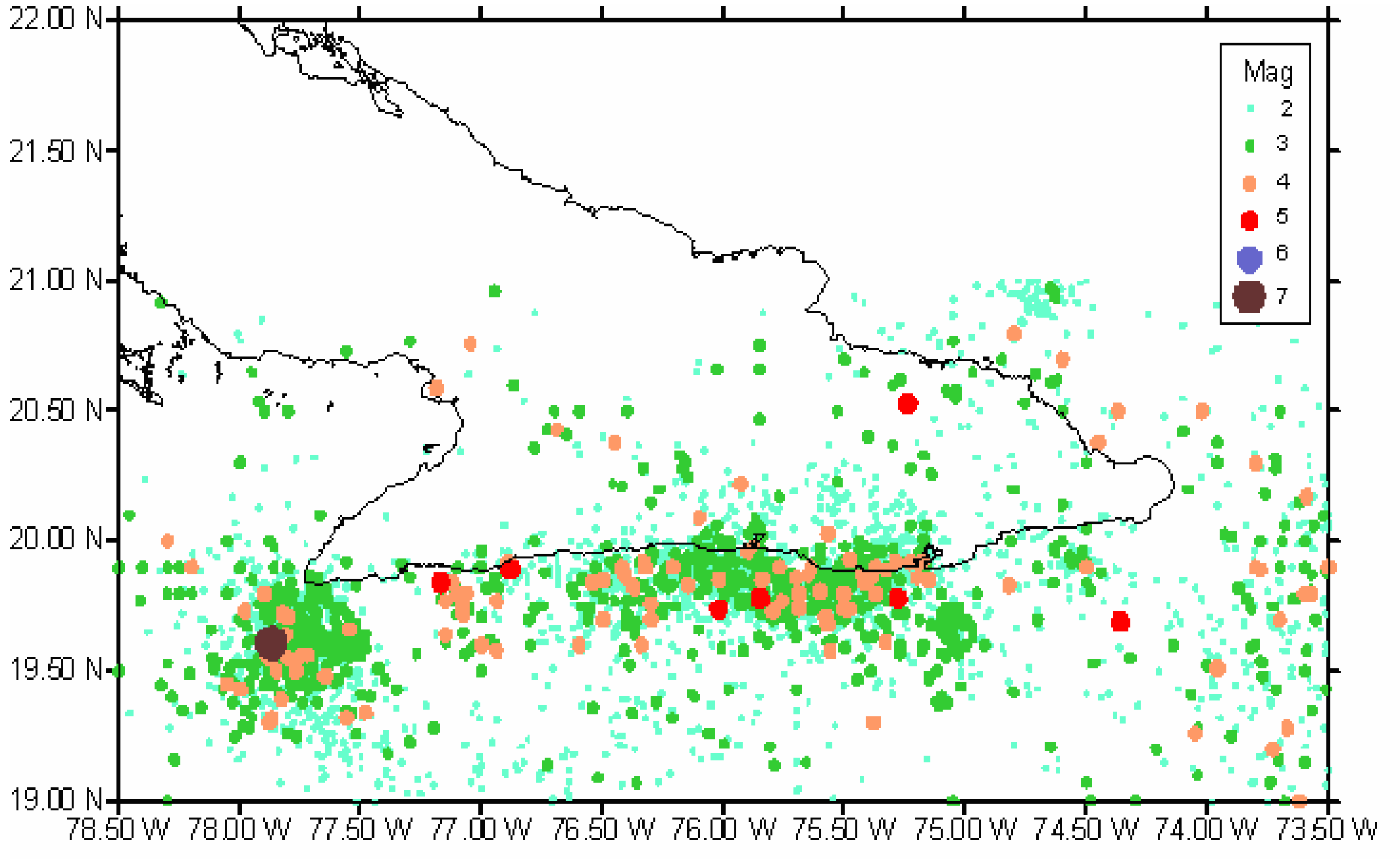

Fig 1. - Seismic region of the eastern Cuba showing epicentres. The magnitude is represented through colour and size of the dots as show in the figure.

An event cannot have a magnitude larger than that of the previous one to be classified as replica. This way, a network of events was obtained (fig 2). An absolute classification of events in main shocks and aftershocks cannot be imposed since in this method a given seism can be simultaneously an aftershock of previous events and a generator of new ones

In the following table the main characteristics of the catalogues of earthquakes is exposed:

| Catalogue | Period | # events | Latitude | Longitude | $m_<$ |
|---|---|---|---|---|---|
| Cuba | 1964 2000 | >6900 | 18.0 - 21.5 | -72.0 - -79.0 | 1.5 |

In our analysis we only consider aftershocks with magnitude smaller than that of the corresponding mainshocks. We consider the critical value $n_c = 10^{-4}$ producing a fragmented network to analyze the diffusion of epicentres in the resulting clusters.
In the determination of the correlation among events the singularities were avoided by limiting the minimum time interval to 60 sec. and the spatial resolution to 100 m.
The distance between the events was determined transforming the latitude and longitude to plane coordinates as in [5].
A scheme of the obtained network is shown is figure 2. Observe the grouping in preferred clusters determining MACs.

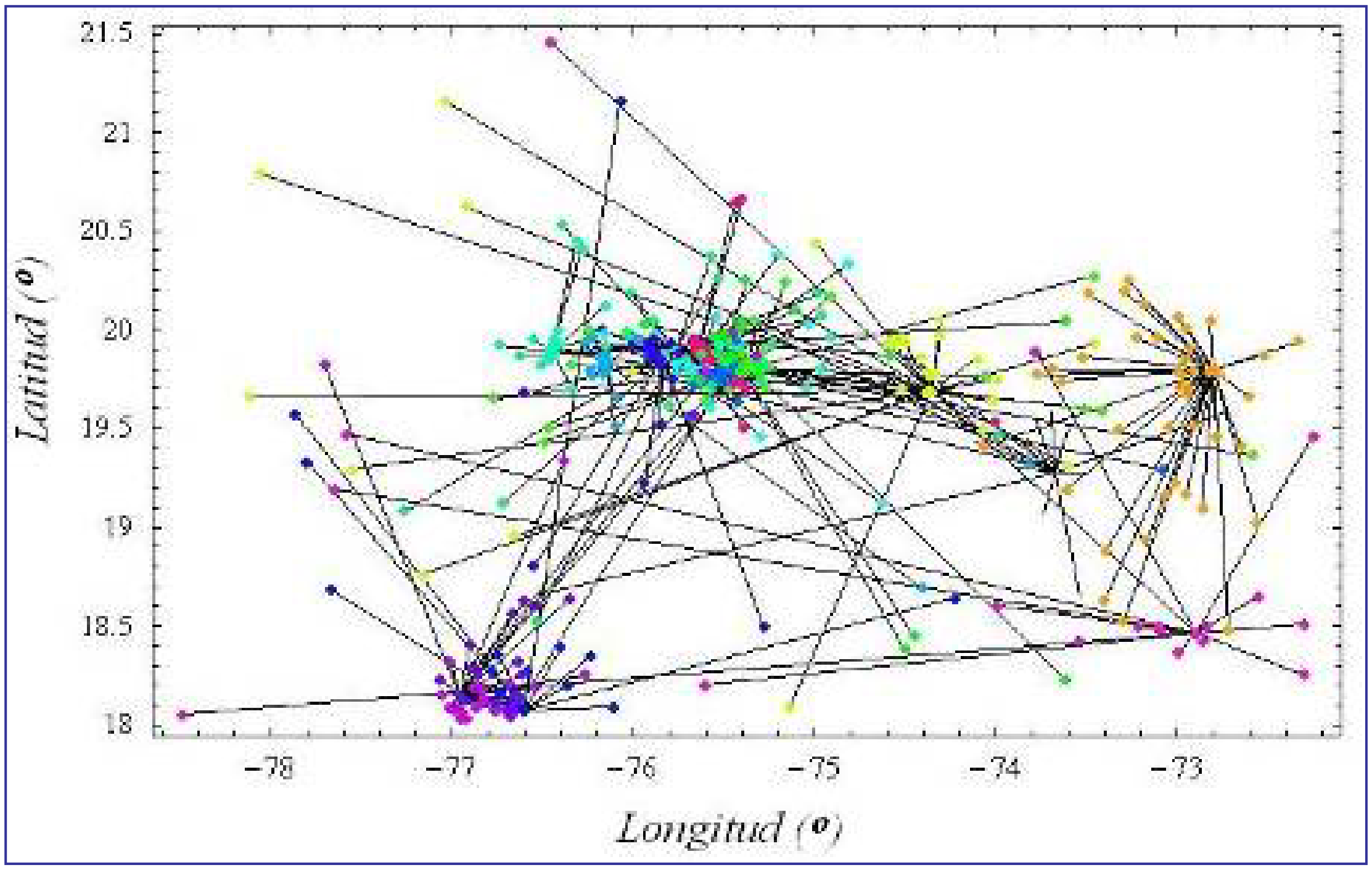

Fig. 2. - Hierarchical network obtained applying the BP method to our catalogue.

## IV.-Results

As already said, with the BP method a network of events is obtained. This network evolves in time in hierarchically organized form where many clusters emerge. There, the aftershocks are linked with their most correlated predecessor. The classification of the events in main shocks and aftershocks is relative because there are many aftershocks with other aftershocks attached as successors.

Then we have a network with the necessary ingredients to consider the scale-free network (SFN) model: The growth by addition of new nodes and the preferential attachment because new nodes connect to a node depending on the degree of the previous one.

The analysis of our network gave a SFN network where the number $k$ of terminals per node is distributed according to $P(k) \sim k^{-\delta}$ with $\delta = 1.84 \pm 0.14$.

With the BP method we obtained 25 MAC containing $N \geq 20$ aftershocks. If, following the catalogue, the variation in the position of each epicentre in any given MAC is studied as a function of time considering the positions of the successive events as the positions of a random walker, such diffusion of the successive positions of the epicentres exhibits a waiting time distribution and the existence of possible obstacles and traps that delay the jumps and memory effects can be inferred.

But, contrary to usual expectation, $P(x,t)$ for different times tends to a peaked distribution around the origin with a tendency to accumulate, not to diffuse.

Figure 3 shows that $< x^2 >$ does not increase with time, but shows dependence as $< x^2 > \sim t^{\alpha}$ with $\alpha < 0$. Time is measured in units of the largest time.

This seems in no way a process of diffusion but of concentration since the mean square displacement decreases with time.

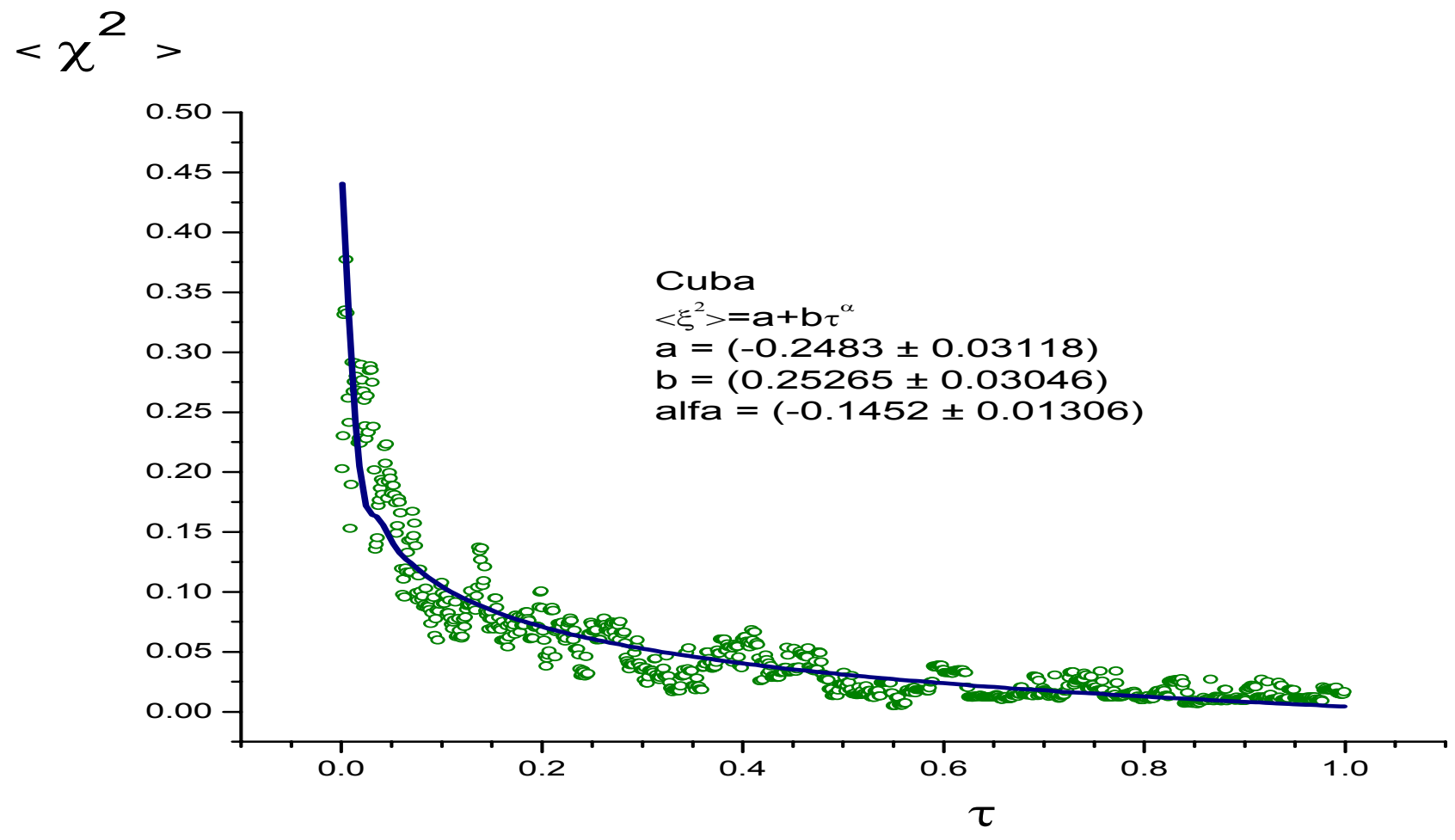

Fig.3. - Mean square displacement as a function of time for the events under consideration, averaged on the 25 MACs obtained. The behaviour is not a normal diffusion (what would be if $\alpha = 1$) neither sub diffusive ($\alpha < 1$) or super diffusive ($\alpha > 1$), but is more like "anti diffusive" since the fitting exponent $\alpha < 0$.

One conclusion could be that there is no diffusion at all, at least among the correlated events, since the objective fact is that correlated earthquakes tend to grouping around the main shock.
Nevertheless, an alternative way of looking at the problem is to analyze if, given that an earthquake modifies its environment in such a way that the region of the epicentre becomes a very fragmented and weakened one, the rest of the stresses in the MAC relax towards the preferential point marked by the main shock, confirming the common assertion of geophysicists about earthquake occurrence: "If it happened, it will happen".

On the other hand, defining $y = 1 - x$ and observing the behaviour of the $y$ dependence of the events with time, it looks like a sub diffusive behaviour, as if the earthquakes diffuse from the frontier of the MAC to the place of occurrence of the main shock. Then it looks that the main shock acts as an attractor of the replica (figure 4)
The correlation among events, the complexity of the environment inside seismic regions and the presence of scaling laws in many of the characteristics of seismic processes lead to think that, if a representation of the distribution of earthquakes in terms of diffusion were possible, a good candidate to represent it is anomalous diffusion, for which fractional dynamics is an adequate mathematical tool.

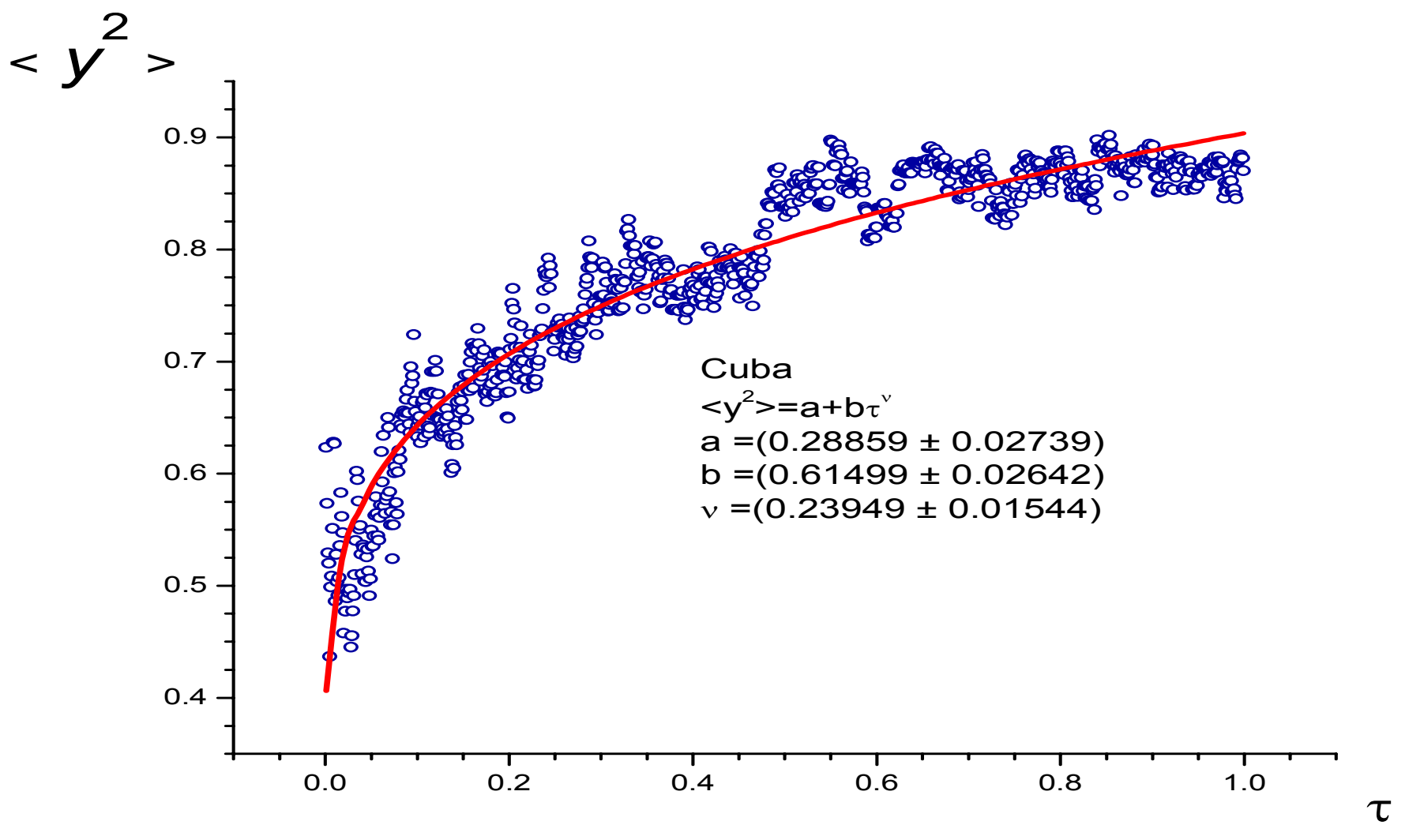

Fig.4. - Behaviour of $y = 1 - x$ with time. This is a CTRW showing that $\langle y \rangle^2 \sim a + bt^{\gamma}$ with $\gamma < 1$, i.e., sub diffusive.

## *V. - Anomalous diffusion*

It is known that the normal diffusion of a particle moving in one dimension can be described by the linear Fokker-Planck equation:

$$\frac{\partial P(x,t)}{\partial t} = K\left[\frac{\partial^2}{\partial x^2} - \frac{\partial}{\partial x}\frac{F(x)}{k_B T}\right]P(x,t)\,, \qquad (1)$$

where $P(x,t)$ is the probability that the particle is at a distance $x$ at time $t$, $K$ is the diffusion coefficient, $F(x)$ the external force, $T$ the temperature and $k_B$ is Boltzmann´s constant.

This equation is usually applied to various types of normal markovian diffusive phenomena. As known, a Gaussian distribution is obtained when the external force $F(x) = 0$. Nevertheless, this equation cannot be applied to the anomalous diffusion phenomenon which is strongly connected with interactions in a non-homogeneous and complex background or with memory effects [9].

To deal with this, a fractional Fokker Planck equation (FFPE) can be introduced (See [11]). Fractional calculus is based in the well-known Cauchy formula:

$$\int_0^t \int_0^{\tau_n} \ldots \int_0^{\tau_{n-1}} f(\tau_n)\, d\tau_n \ldots d\tau_1 = \frac{1}{(n-1)!}\int_0^t (t-\tau)^{n-1} f(\tau)\, d\tau \qquad (2)$$

On this basis the fractional integral is defined as a generalization of this expression via [10-12]:

$$ {}_aD_t^{-\alpha} f(t) = \frac{1}{\Gamma(\alpha)} \int_a^t (t-\tau)^{\alpha-1} f(\tau)\, d\tau \qquad (3)$$

and consequently the fractional derivative can be defined through fractional integration and successive ordinary differentiation as:

$$\frac{\partial P(x,t)}{\partial t} = {}_0D_t^{1-\alpha}\left[ K_\alpha \frac{\partial^2}{\partial x^2} P(x,t)\right],$$

where the symbolics:

$$ {}_aD_t^{\alpha} f(t) = \begin{cases} \dfrac{d^m}{dt^m}\left[\dfrac{1}{\Gamma(m-a)} \displaystyle\int_a^t \dfrac{f(\tau)}{(t-\tau)^{\alpha-m+1}} d\tau\right] & \text{m-1<}\alpha\text{<m} \\ \dfrac{d^m}{dt^m} f(t) \quad \alpha\text{=m} \end{cases} \qquad (4)$$

has been introduced.

This definition of fractional derivative involves not only the time of evaluation of the derivative, but also all previous times. This makes fractional derivative useful to account processes whose properties depend on their history.
In this formulation the fractional diffusion equation reads:

$$\frac{\partial P(x,t)}{\partial t} = {}_0D_t^{1-\alpha}\left[ K_\alpha \frac{\partial^2}{\partial x^2} P(x,t)\right]. \qquad (5)$$

The solution of this equation can be found as [10]:

$$P(x,t) = \frac{1}{\alpha}\left(\frac{t}{|x|^{1+\frac{2}{\alpha}}}\right) L_{\frac{\alpha}{2}}\left(\frac{t}{|x|^{\frac{2}{\alpha}}}\right) \qquad (6)$$

$L_{\frac{\alpha}{2}}(z)$ is a one sided Lévy stable density [13] which for small $\alpha$ can be approximated as (See [10]):

$$P(x,t) \sim \frac{1}{\alpha\, t^{\alpha/2}} \exp\left(-c \frac{|x|}{t^{\alpha/2}}\right), \qquad (7)$$

where $c$ is a constant dependent on $\alpha$. This solution is represented in Fig.5 for arbitrary values of $\alpha$ and $c$. Notice the difference with the Gaussian distribution corresponding to normal diffusion. Equation (6) expresses an important property of the anomalous diffusion: it follows Lévy statistics.

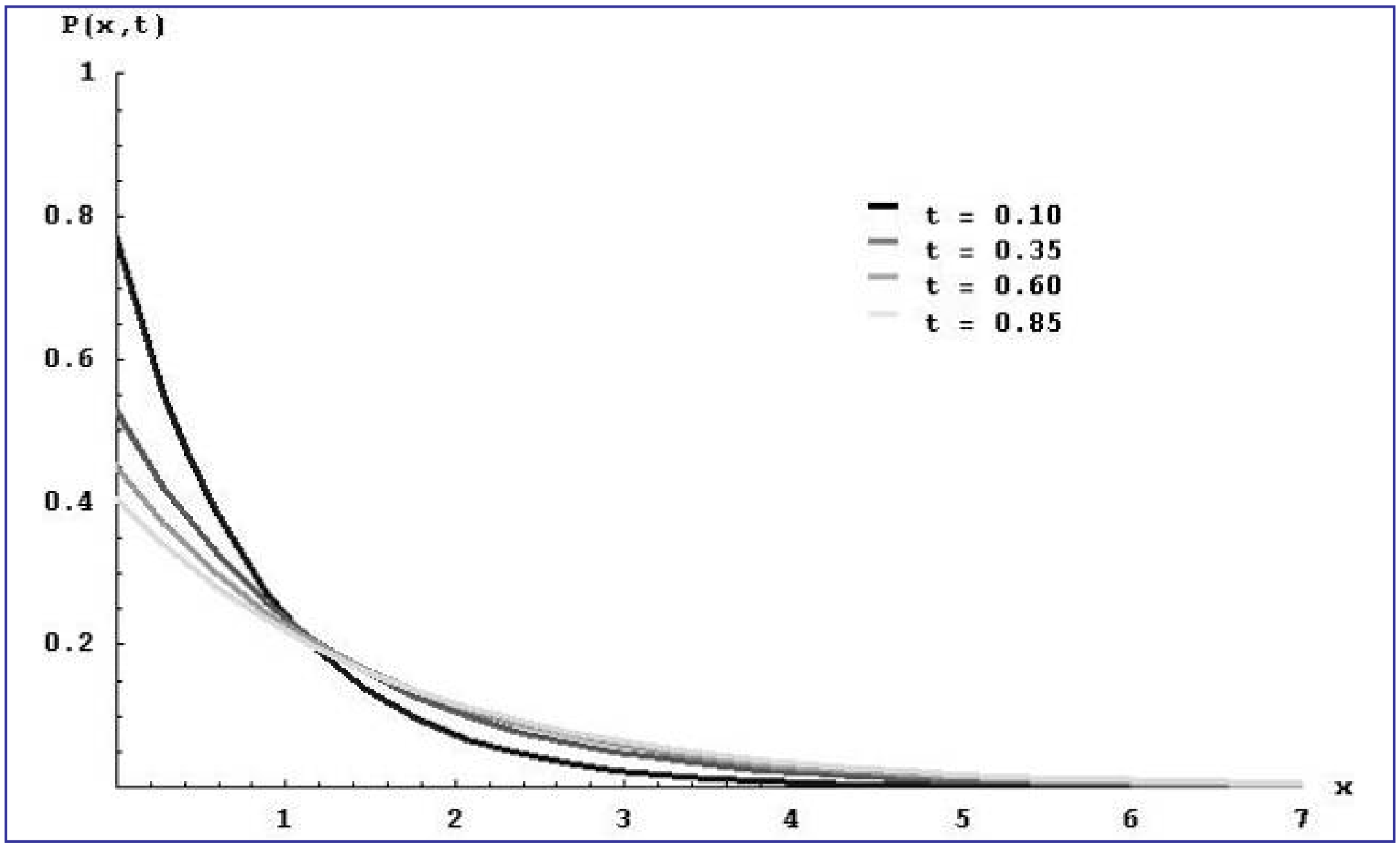

Fig. 5. - Representation of equation (7) for different values of t., $c = 1$, and $\alpha = 0.1$. For diffusion governed by CTRW the solution displays a cusp at $x = 0$

## *VI. – Discussion and Conclusions*

The application of the BP method has revealed not only the possibility to group earthquakes in complex SFN networks and an objective and more quantitative classification of earthquakes in main shocks and aftershocks. The introduction of a diffusive description for the positions of the epicentres into the spatiotemporal cells generated by this method exhibits an abnormal diffusive behaviour (antidiffusion).

The apparent process of "antidiffusion" seems to be, in our interpretation, the result of taking main shock-aftershock cells of finite extension and try to describe spatial distribution of earthquakes as the migration of one event inside the cell.

Indeed, if we assume that Equation (6) is valid as a solution of a FFPE in the MAC or at least assume that the most significant values of the distribution occur there, then, to calculate $< x^2 >$ we may take finite limits of integration. Then if we put the approximate expression:

$$< x^2 > \approx \int_0^1 x^2 P(x,t)\,dx\,, \qquad (8)$$

the values of the distribution inside a finite region are being used to calculate the mean square displacement inside such a region.

Besides, in the calculation (8) we take for expression (6) the asymptotic form $L_{\alpha/2}\left(\frac{t}{x^{2/\alpha}}\right) \sim \left(\frac{t}{x^{2/\alpha}}\right)^{-\alpha/2-1}$ of Levy distributions for large values of the arguments. As the earthquakes are limited to occur in a finite region, this means that the values of the displacement are smaller than those corresponding to a true diffusion process. This gives behaviour like $< x^2 > \sim t^{-\alpha/2}$.

Then, the antidiffusive behaviour here observed can be attributed to a sort of "size effect".
This effect reveals the most correlated events as acting like an ensemble determining a whole collective process of energy relaxation in the area, so that each earthquake in the MAC is just part of the process and is linked to the others in a cooperative way, where the main shock triggers the energy liberation preferentially from the frontier of the region that occupy the most correlated earthquakes to the point where the main event occurred.

So, epicentre diffusion in MACs determined by the BP method can be described through CTRW behaviour, from the frontier of the MAC to the main shock, confirming the omnipresence of Lévy processes in many natural phenomena, and presenting the main shock as some kind of "attractor" of replicas.

The main point is that this is not a characteristic of the catalogue we have used here. This catalogue was presented only as a tool to exemplify the curious diffusive behaviour that the BP method reveals in earthquakes. Our conclusions have been tested also with catalogues of earthquakes from other places like Spain, Greece and California with essentially the same results, and a more extensive report will be published elsewhere.

Besides, the BP method should not be blamed for these results. Some artificial data of different nature were generated, to which we applied the BP method. Similarly to the Cuban catalogue, MACs with more than 20 events were taken into account. The catalogues were of different types:
a) Events occurring at uniform time steps, with Brownian motion and magnitudes distributed according to the Gutenberg-Richter law: this catalogue exhibits normal diffusion in MACs: $< x^2 > \sim t$ .
b) Events Levy-distributed in space but with magnitudes uniformly distributed: this exhibits sub-diffusive behaviour, as expected.
c) Data with random uniform distribution of both location and magnitudes: show no clear diffusive behaviour.

So, antidiffusion of epicentres looks an effect linked with the correlation between seisms and the consideration of their diffusion in a limited space-time region.